%% file: main.tex
\documentclass[twocolumn]{aastex63}

\usepackage[utf8]{inputenc}
\usepackage{natbib}
\usepackage{graphicx}
\usepackage{amsmath}
\usepackage{booktabs}
\usepackage{longtable}
\usepackage{xspace}
\usepackage{hyperref}
\usepackage[T1]{fontenc}
\usepackage{lipsum}
\usepackage{comment}
\usepackage{xcolor}
\usepackage{enumitem}
\usepackage{multirow}
\usepackage{IEEEtrantools}
\usepackage{rotating}

\def\intra{\mathrm{intra}}
\def\slot{\mathrm{slot}}

\def\multivisit{\mathrm{multi-visit}}
\def\singlevisit{\mathrm{single-visit}}

\def\Acal{\mathcal{A}}
\def\Dcal{\mathcal{D}}

\def\Rcal{\mathcal{R}}
\def\Scal{\mathcal{S}}
\def\Tcal{\mathcal{T}}

\newcommand{\AstroQ}{\textit{AstroQ}\xspace}

\newcommand{\isr}[2]{\ensuremath{#1_{\rm #2}}\xspace}

\usepackage{xparse} 
\usepackage{xcolor} 
\usepackage{etoolbox} 
\input{params}

\begin{document}

\title{AstroQ: Automated Scheduling of Cadenced Astronomical Observations}

\author[0000-0001-8342-7736]{Jack Lubin}
\affiliation{Department of Physics \& Astronomy, University of California Los Angeles, Los Angeles, CA 90095, USA}

\author[0000-0003-0967-2893]{Erik A. Petigura}
\affiliation{Department of Physics \& Astronomy, University of California Los Angeles, Los Angeles, CA 90095, USA}

\author[[0000-0002-8952-5617]{Velibor V. Mi\v{s}i\'{c}}
\affiliation{Anderson School of Management, University of California Los Angeles, Los Angeles, CA 90095, USA}

\author[0000-0002-4290-6826]{Judah Van Zandt}
\affiliation{Department of Physics \& Astronomy, University of California Los Angeles, Los Angeles, CA 90095, USA}

\author[0000-0002-9305-5101]{Luke B. Handley}
\affiliation{Cahill Center for Astronomy $\&$ Astrophysics, California Institute of Technology, Pasadena, CA 91125, USA}

\begin{abstract}
Astronomy relies heavily on time domain observations. To maximize the scientific yield of such observations, astronomers must carefully match the observational cadence to the phenomena of interest. This presents significant scheduling challenges for observatories with multiple large programs, each with different cadence needs. To address this challenge, we developed \AstroQ, an automated framework for scheduling cadenced observations. We tested this on a suite of Doppler exoplanet programs at Keck Observatory, where the algorithm powers the KPF-Community Cadence project. As a point of reference, \AstroQ can determine the provably optimal ordering of 3680 observations of 200 targets --- each with its own cadence needs and accessibility constraints --- over a six month period to five minute time resolution. Schedules of this size may be constructed in $\sim$120 seconds on modern workstation, enabling dynamic rescheduling due to weather changes, target-of-opportunity interrupts, and other needs. A key advantage of \AstroQ over manual scheduling is realistic projections of program completion, savings in human effort, and elimination of human bias in balancing many programs. \AstroQ is open source and may be applied to other scheduling needs, both in astronomy and beyond. 
\end{abstract}
\keywords{scheduling optimization, radial velocities}

\section{Introduction}
\label{Intro}
Many of our most profound insights into the secrets of the Universe have been painstakingly won by astronomers working in the time domain. There are too many examples to count, but they range from the precession Mercury's orbit (established over centuries) that pointed toward general relativity \citep{LeVerrier59}, to the Cepheid period-luminosity relationship (observable over weeks) that set the scale of the Milky Way and Local Group \citep{Leavitt12}, to the tiny Doppler wobbles observed in some stars (periods ranging from days to decades) that reveal companion planets \citep{Mayor95}. Observational campaigns to monitor time-variable phenomena require fine control of observational cadence --- the {\em timing of} and {\em spacing between} observations. This is particularly challenging at the world's largest telescopes where a large number of instruments and programs competitively vie for limited time. 

In this work, we develop, implement, and test an algorithm that schedules observations from a suite of programs with their own cadence requirements and accessibility constraints. We refer to the broad class of automatic scheduling algorithms as `autoschedulers' and our particular implementation as {\em AstroQ}. 

Our work here is motivated by our needs as exoplanet observers who use precise radial velocity (PRV) observations of stars to detect the reflex motion of planets. In this field, one must specify and control observing cadence to adequately sample of orbital phase and to disentangle stellar and planetary signatures. For motivation on the role of cadence in PRV science, see \citet{kpfcc}, \citet{Handley2024_Semester}, and references therein. 

\AstroQ was also motivated by a specific need at W.~M.~Keck Observatory. Historically, PRV programs at Keck were executed in an {\em ad hoc} queue that needed to schedule some $\sim$3000 observations per six-month semester. Whatever the objective function, with $\sim$3000! possible schedules a human cannot find an optimal schedule. Manually generating observing schedules is labor intensive and required roughly 1.0 Full Time Equivalent (FTE) which is  typical for PRV queues \citep{kpfcc}. Human-generated schedules naturally lead to questions regarding fairness when balancing programs. It is also impossible to reliably forecast completion rates or factor in weather losses. 

The first PRV autoscheduler deployed at Keck is described in \citet{Handley2024_Semester}. In brief, this algorithm takes a request set flowing from different programs and does an initial coarse scheduling by assigning observations to quarter-night bins, while respecting the specified inter-night cadence (the minimum spacing between observations on different nights). The coarse schedule for the upcoming night is refined by solving the `Traveling Telescope Problem' \cite{Handley2024_TTP} to determine the optimal ordering of observations. 

While this first algorithm could successfully schedule an entire semester, it struggled in a number of ways we wish to address here: (1) there were often significant gaps between the quarters that had to be filled in manually, (2) it struggled to gracefully handle concentrations or gaps in right ascension (RA) coverage, (3) it did not support intra-night cadence (the minimum spacing between observations on the same night), and (4) it did not naturally handle custom observational strategies such as observations timed for a particular orbital phases. 

This work is organized as follows: In \S\ref{sec:algorithm} we describe the integer linear programming formulation of the scheduling problem. Then, in \S\ref{sec:keck}, we explain how \AstroQ is implemented at Keck Observatory as part of the Doppler planet searches conducted with the Keck Planet Finder (KPF, \citealt{Gibson24}). In \S\ref{sec:performance}, we benchmark this code on a realistic request set and note program completion rates and computational complexity. We offer some concluding thoughts and ideas for future improvements in \S\ref{sec:conclusion}. Before departing this introduction, we note that while \AstroQ was motivated by the exoplanet community, it is a general-purpose code, applicable to a wide variety of science cases. To facilitate the widest possible usage, we have made our Python code open source and publicly available.%
\footnote{\url{https://github.com/jluby127/AstroQ}} 

\section{AutoScheduler}
\label{sec:algorithm}

\begin{figure*}[t!]
\centering\includegraphics[width=1\textwidth]{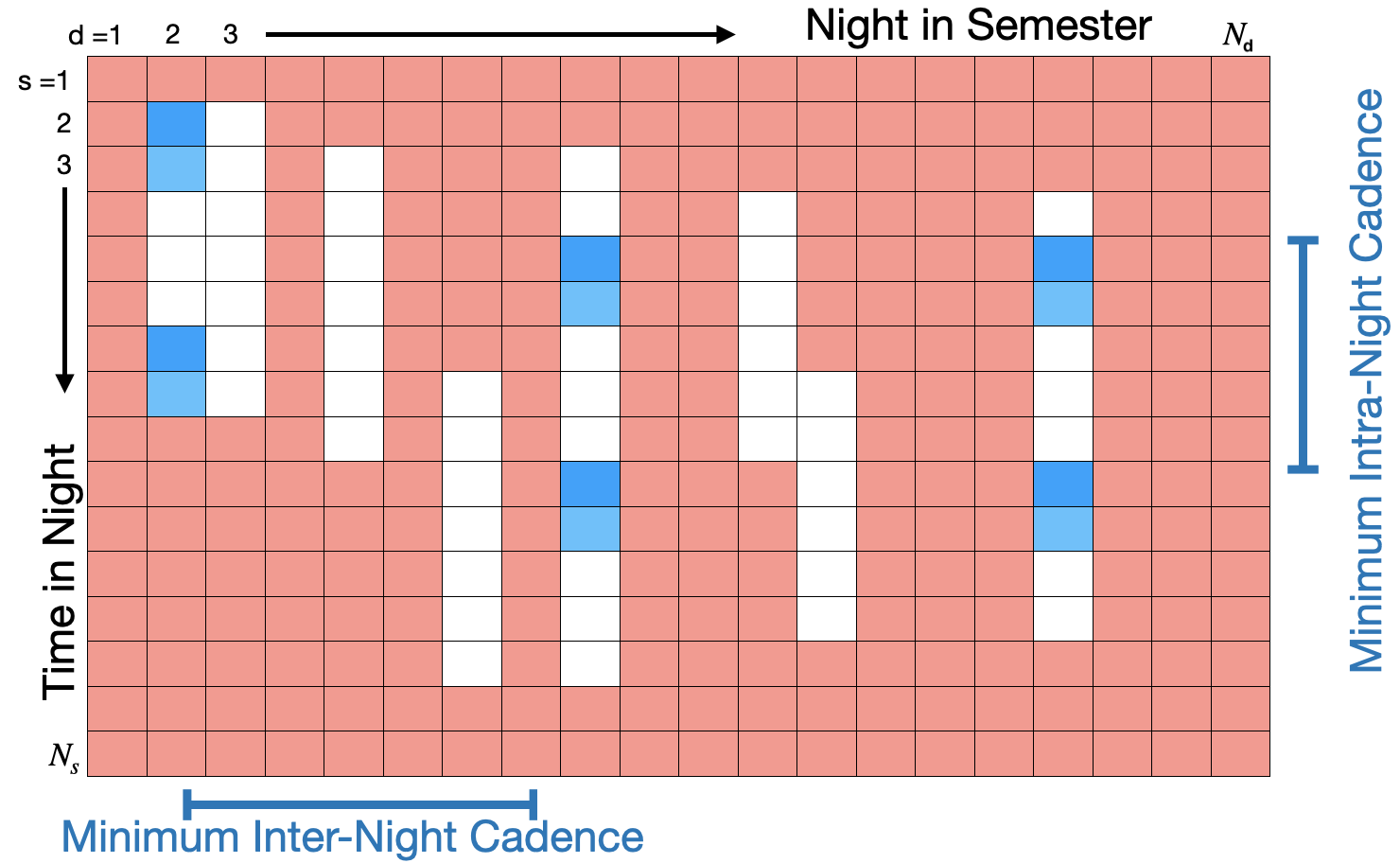}
\caption{{\bf Day-slot scheduling framework.} \AstroQ discretizes time in an observing semester into $N_d \times N_s$ slots of equal length, where $d$ and $s$ labels each day and slot. Each request is accessible for a subset of all possible $(d,s)$ pairs, and the inaccessible slots are red. The binary decision variable $Y_{r,d,s}$ is 1 if a visit of request $r$ begins in slot $s$ on day $d$ (shown as dark blue squares) and 0 otherwise. Here, the request specifies \isr{t}{visit,r} = 2 slots to complete a visit. The light blue cells indicate the $(d,s)$ pairs where no other observations may commence. The request specifies a minimum {\em inter-night} cadence \isr{\tau}{inter,r} of 6 days, i.e. the  minimum number of days between successive visits. It also species a minimum {\em intra-night} cadence \isr{\tau}{intra,r} of 3 slots.}
\label{fig:slot-grid}
\end{figure*}

\begin{figure*}[t!]
\centering\includegraphics[width=1\textwidth]{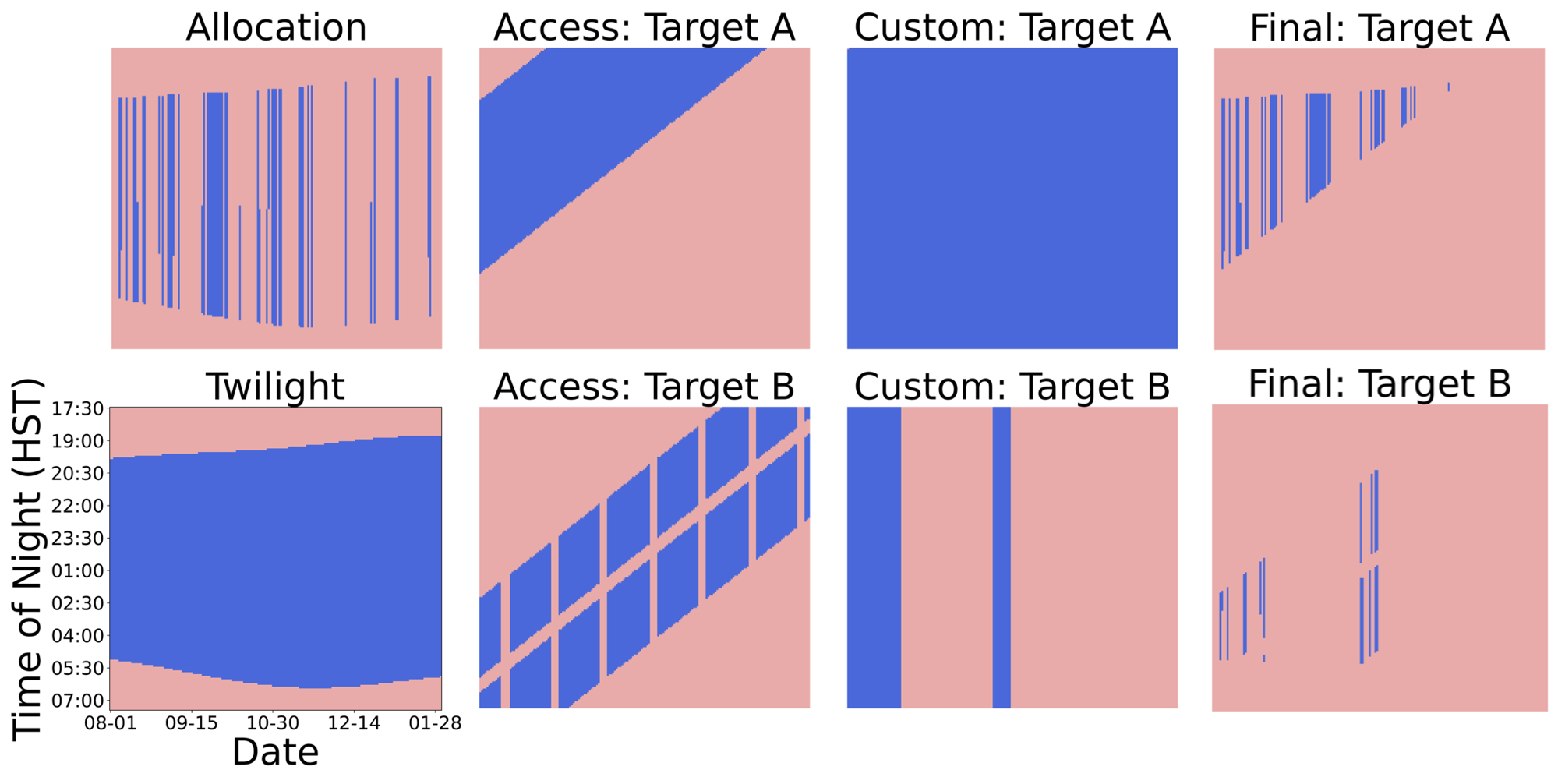}
\caption{{\bf Construction of two accessibility maps.} Each panel is analogous to Figure~\ref{fig:slot-grid} and share the same limits. The left column shows the Allocation and Twilight maps drawn from the Keck 2018B semester (see \S\ref{sec:2018B}), which are the same for both requests. The two requests have different coordinates and thus different accessibility maps. For Target A, the upward tilt of the blue band indicates progressively later rise/set times over a semester. For Target B, the vertical stripes are due to moon `collisions' and the diagonal exclusion region is due to zenith crossing events. Target A has no custom accessibility constraints, while Target B is restricted to a range of dates. The right column shows the final maps $A_r$, which are the intersection of all constituent maps. Each request's final accessibility set $\Acal_r$ consists of the blue $(d,s)$ pairs.} 
\label{fig:accessibility-maps}
\end{figure*}

\subsection{Scheduling Framework} 

\AstroQ takes as input $N_r$ observing requests. These requests instruct the telescope to `visit' (i.e. slew to, acquire, and observe) a specific target subject to accessibility and timing constraints explained below. We use the word `visit' as the fundamental scheduling quantum as opposed to `observation' or `exposure'. A single visit may contain multiple observations or exposures, but these details are irrelevant for scheduling purposes. Visits flowing from these requests are scheduled over an interval or `semester' spanning $N_d$ days and each day is divided into $N_s$ discrete slots with equal duration $\tau_{\slot}$. \AstroQ schedules requests into one or more of these slots.%
\footnote{This work builds heavily upon \citet{Handley2024_Semester} and adopt their terminology where possible. However, the definition of slot is different in the two works}
Figure~\ref{fig:slot-grid} shows this framework graphically. Throughout this work, we refer to an individual request as $r$ and the set of all requests as $\Rcal$. We follow a similar convention for days $d \in \Dcal$ and slots $s \in \Scal$. For a full list of the definitions for all variables and values used in this work, see Appendix Table~\ref{variableDict}.

Each request $r$ contains the following information specifying the observational strategy and target accessibility:

\begin{itemize}
    \item {\em Number of nights}, \isr{n}{inter,r} [integer]. {\em Maximum} number of unique nights to visit target.

    \item {\em Inter-night spacing}, \isr{\tau}{inter,r} [integer]. The {\em Minimum} number of days that must elapse before a target may be scheduled again. When \isr{n}{inter,r} = 1, we set \isr{\tau}{inter,r} = 0.

    \item {\em Maximum number of visits within a night}, \isr{n}{intra,max,r} [integer]. A PI may request multiple visits within a night. 

    \item {\em Minimum number of visits within a night}, \isr{n}{intra, min, r} [integer]. For certain science cases there may be a minimum number of visits to be scientifically useful (e.g., activity monitoring). For these cases, a PI may set a minimum acceptable number of visits. 
    
    \item {\em Intra-night spacing}, $\tau_{\intra,r}$ [integer]. The {\em minimum} time, in slots, between visits of the same target on the same night. When \isr{n}{intra, min, r} = 1, we set \isr{\tau}{intra, r} = 0.
    
   \item {\em Number of slots per visit}, \isr{t}{visit,r} [integer], representing how many consecutive slots a single visit requires to complete the observation.

    \item {\em Set of accessible slots}, $\mathcal{A}_{r}$ [tuples]. The set of $(d,s)$ pairs where $r$ is accessible. In set-builder notation (used throughout this paper) this is: 
    \begin{align*}
    \mathcal{A}_{r} = \{(d,s) \in \Dcal \times \Scal \mid &\text{request $r$ is accessible}  \\ 
    &\text{in slot $s$ of day $d$} \}.
    \end{align*}

    We describe how we compute $\mathcal{A}_{r}$ for the KPF Community Cadence project in \S\ref{subsec:unavailable} but this procedure will be different for every deployment of \AstroQ.

\end{itemize}

\subsection{Decision Variables}
\label{variables}

Decision variables include:

\begin{itemize}

\item $Y_{r, d, s}$ -- the `yes' matrix. $Y_{r, d, s}$ = 1 if request $r$ is scheduled to begin in slot $s$ on night $d$, and 0 otherwise. While one may conceptualize $Y_{r, d, s}$ as a 3D array, in practice it is implemented as a 1D array indexed by the 3-tuple $(r, d, s) \in \Acal$, the set of all valid requests, days, and slots. Since only a fraction of slots are available to each target, this sparse representation significantly reduces the model size compared to a dense, rectilinear construction.

\item $W_{r,d}$ -- the `day-tracking' matrix. $W_{r,d}$ = 1 if request $r$ is scheduled to any slot $s$ within night $d$, and 0 otherwise. It is used to enforce intra-night cadence constraints. $W_{r,d}$ is only defined for multi-visit requests, i.e. $r \in \Rcal_{\multivisit}$ where $\isr{n}{intra,min,r} > 1$. 

\end{itemize}

\subsection{Objective Function}
\label{sec:objective}

We seek to maximize the time-weighted completion fraction of all requests by minimizing:

\begin{equation}
    \mathrm{obj} = \ \sum_{r\in \Rcal} \Theta_r \isr{t}{visit,r}
\label{objective_function}
\end{equation}
where $\Theta_r$ is the `shortfall' matrix that tracks the number of unscheduled observations of request $r$ over the full semester and is  defined via 
\begin{align}
\Theta_r & = \max \{ 0, a_r \}, \label{eqn:theta}\\
a_r & =\isr{n}{inter,r} - P_r - \big(\frac{\sum_{ (d,s) \in \mathcal{A}_r} Y_{r,d,s}}{\isr{n}{intra, max, r}}\big). \label{eqn:a}
\end{align}
Here, $P_r$ is the number of prior visits. Equation~\ref{eqn:a} is defined so that $a_r$ equals zero when the sum of the number of past and the number of future planned observations equals $\isr{n}{inter,r}$, i.e. when the request is complete. Multi-visit requests that do not achieve \isr{n}{intra,max,r} result in fractional $a_r$. While the auxiliary variable $a_r$ may be negative by exceeding \isr{n}{inter,r}, $\Theta_r$ may not; thus, the objective function does not reward extra visits. By defining $\Theta_r$ in terms of a maximum, we create a convex piecewise linear objective function that is constructed using linear constraints and auxiliary variables.

\subsection{Constraints}
\label{sec:constraints}

\newcounter{constraint}
\setcounter{constraint}{1}
\label{const:prevent-overlapping}
\noindent\textbf{Constraint \arabic{constraint}: Prevent overlapping visits.} If all visits required a single slot, we could prevent overlapping visits via:

\begin{equation}
\sum_{r \in \Rcal_{d,s}} Y_{r,d,s} \leq 1, \quad \forall\ (d,s) \in \Tcal, 
\label{eqn:const1}
\end{equation}
where $\Tcal$ is the set of all $(d,s)$ pairs where at least one $r$ is accessible. However, to ensure no multi-slot visits overlap, we require that, in order for any new observation to begin at $(d,s)$, no previous observation may begin within \isr{t}{visit} slots of $(d,s)$:
\begin{align}
\sum_{r \in \Rcal_{d,s}} Y_{r,d,s} 
&\leq 1- 
\sum_{\delta = 1}^{s - 1} 
\sum_{r' \in \Rcal^{\prime}_{d,s,\delta}} 
Y_{r',d,s-\delta}, \\
&\forall (d, s) \in \Tcal \nonumber.
\label{eqn:prevent-overlapping-visits}
\end{align}
Here,
\begin{align}
\Rcal^\prime_{d,s,\delta} = \{ r \in \Rcal_{d,s} | \isr{t}{visit,r} \geq \delta \}
\end{align}
is the set of requests that, if they were scheduled $\delta$ slots before $s$, would overlap with $(d,s)$. Note that Equation~\ref{eqn:const1} is always satisfied by Equation~\ref{eqn:prevent-overlapping-visits}, so we may simply include Equation~\ref{eqn:prevent-overlapping-visits} without loss of generality.

\refstepcounter{constraint}
\label{const:max-visits}
\noindent\textbf{Constraint \arabic{constraint}: Enforce maximum number of visits.} For single-visit requests, we require the sum of the future scheduled visits and past visits not exceed the maximum value:

\begin{equation}
\begin{aligned}
\sum_{(d,s) \in \mathcal{A}_r}Y_{r,d,s} + P_r \le \isr{n}{inter,r} \quad \forall\ r \in \Rcal_{\singlevisit}.
\label{eqn:const2a}
\end{aligned}
\end{equation}
For multi-visits, we replace the left hand side with $W_{r,d}$ which is 1 regardless of how many times an observation was scheduled on $d$:

\begin{equation}
\begin{aligned}
\sum_{d \in \Dcal_r}W_{r,d} + P_r \le \isr{n}{inter,r}\quad \forall\ r \in \Rcal_{\multivisit}.
\label{eqn:const2b}
\end{aligned}
\end{equation}

\refstepcounter{constraint}
\label{const:inter-night}
\noindent\textbf{Constraint \arabic{constraint}: Enforce inter-night cadence.} If a request $r$ is scheduled on day $d$ and slot $s$ all $(d,s)$ pairs within \isr{\tau}{inter,r} of $d$ must be excluded. This is achieved by:
\begin{align}
\sum_{s \in \Scal_{r,d}} \frac{1}{\isr{n}{intra,max,r}} Y_{r, d, s} & \leq  
1 - \sum_{\delta = 1}^{\isr{\tau}{inter, r} -1} \sum_{s \in \Scal_{r,d+\delta}} Y_{r, d + \delta, s}, \\
& \forall\ (r,d) \in \Rcal \times \Dcal .
\label{eqn:const4}
\end{align}

\refstepcounter{constraint}
\label{const:intra-night}
\noindent\textbf{Constraint \arabic{constraint}: Enforce intra-night cadence.} For multi-visit requests, we must ensure that at least \isr{\tau}{intra,r} slots elapse between repeat visits on a given night. This is achieved by:


\begin{align}
    Y_{r,d,s} & \le W_{r,d} - \sum_{s^\prime \in \mathcal{S}^{\prime}_{r,d,s} } Y_{r, d, s^\prime} \label{eqn:const5} \\
    &\forall\ \{(r,d,s) \in \Acal | r \in \Rcal_{\multivisit} \} \notag
\end{align}

where
\[
\mathcal{S}^\prime_{r,d,s}  = \{s^\prime \in \mathcal{S}_{r,d} | s < s^\prime < s + \isr{\tau}{intra,r}\} .
%
\]
Note, we could achieve the same logical constraint by replacing $W_{r,d}$ with 1, but since the constrain is only relevant nights when at least one visit of $r$ is scheduled our formulation achieves a tighter formulation of the relaxed linear program and tends to reduce solve times. 

\refstepcounter{constraint}
\label{const:W}
\noindent\textbf{Constraint \arabic{constraint}: Bound total number of repeat visits.} As part of the request strategy, a PI may select both a minimum and maximum number of visits within a night. While a single value may be specified by setting $\isr{n}{intra,min} = \isr{n}{inter,max}$, for some programs a PI may find it advantageous to allow $\isr{n}{intra,min} < \isr{n}{inter,max}$. A target's nightly accessibility may make $\isr{n}{inter,max}$ visits infeasible. Setting $\isr{n}{intra,min} = \isr{n}{inter,max}$ creates an all-or-nothing scenario; if all visits are not possible, none will be scheduled. Allowing a range of visits makes more nights available and will lead to a greater number of total visits. This is achieved by:

\begin{align}
\sum_{s \in S_{r,d}} Y_{r,d,s} & \geq \isr{n}{intra,min,r}  W_{r,d}, \\
\sum_{s \in S_{r,d}} Y_{r,d,s} & \leq \isr{n}{intra,max,r}  W_{r,d}, \\
& \forall\ r \in \mathcal{R}_{\mathrm{multi-visit}} \notag.
\label{eqn:const6}
\end{align}

\section{Implementation at Keck Observatory}
\label{sec:keck}
The previous section described the mathematical formalism behind \AstroQ, which may be employed at any telescope or even to non-astronomical observations. Here, we describe how our team uses \AstroQ to schedule Doppler observations at Keck Observatory as part of the KPF Community Cadence program. 

\subsection{Defining the day-slot grid}

Telescope time at Keck is awarded in half-year increments: an `A' semester (August 1 to January 31) and a `B' semester (February 1 to July 31). We divide each of the $\approx$180 nights in a semester into $N_s$ = 168 slots that are $\tau_{\slot}=5$~min each. We set the start and end times to 17:30 and 07:30 HST to accommodate the earliest/latest sunset/sunrise times at Maunakea. Our choice of $\tau_{\slot}$ balances time resolution with computational expense. The grid contains a total of $\approx  180 \times 168 \approx 30,000$ possible $(d,s)$ pairs per request.

\subsection{Determining Accessible Slots}
\label{subsec:unavailable}

Each request $r$ is only accessible for a subset, $\mathcal{A}_r$, of all possible $(d,s)$ pairs. A target's nightly accessibility is based on a number of factors, which we list below. It is useful to conceptualize each factor as a binary ``accessibility map'' $A_{k,r}$ with dimensions $N_d \times N_s$ for each factor $k$. For example, if a target passes our altitude/azimuth limits for $(d,s)$ then the $(d,s)$ element of $\isr{A}{alt/az,r} = 1$ and 0 otherwise. 

\begin{itemize}
\item {\em Instrument allocation}. Only certain nights and certain portions of those nights are allocated to KPF Community Cadence.

\item {\em Twilight}. Observations must occur between 12~deg evening/morning twilight times.

\item {\em Altitude/azimuth}. Our altitude limits depend on target declination and telescope azimuth. For all targets, we set a lower limit of 18~deg to avoid extreme airmass and differential atmospheric refraction as well as shutter vignetting; we set an upper limit of 85~deg because the azimuth drive cannot keep up with targets transiting near zenith. Additionally, we account for the Nasmyth platform which excludes altitudes below 33~deg between 5~deg and 146~deg in azimuth. Lastly, for targets that are comfortably between the extremes of Maunakea accessible declinations  ($\delta = -30$~deg to 75~deg), we set a lower limit of Alt > 33~deg to favor higher airmasses and reduce demands of the KPF tip-tilt corrector. We use the vectorized time/coordinate functions in {\em AstroPy} for efficient computation. 

\item {\em Moon}. We require targets to be at least 30~deg from the moon (as determined by {\em AstroPy}).

\item {\em Prior observations}. While Constraint~\ref{const:inter-night} enforces inter-night cadence between any two upcoming visits, we also must also ensure that the first upcoming visit does not occur within \isr{\tau}{inter,r} of a past visit. We query the database of past observations and remove $(d,s)$ that violate \isr{\tau}{inter,r}.

\item {\em Custom accessibility}. A PI may choose to further constrain target accessibility by enforcing a more restrictive minimum elevation limit, hit specific orbital phases, coordinate simultaneous observations with another observatory, or other reasons. As with other constraints, custom accessibility rules do not guarantee that visits will occur in the allowed times; they simply forbid visits outside these times. Overly restrictive timing constraints risk lower program completion. 

\item {\em Multi-slot visits}. We take the intersection of the aforementioned accessibility maps to produce an `instantaneous accessibility map.' This map specifies whether a target is accessible during that $(d,s)$ pair. To schedule the visit, the target must be accessible during all \isr{t}{visit} - 1 slots {\em after} the starting $(d,s)$. We accomplish this by growing the zeroes in the $A_r$ backward by \isr{t}{visit,r} to produce our ultimate map. 

\end{itemize}

A target's final accessibility map is the intersection of these separate maps $A_{r} = \prod_k A_{k,r}$ and is shown schematically in Figure~\ref{fig:accessibility-maps}. \AstroQ uses $A_r$ to construct the various accessibility sets ($\mathcal{A}_r$, $\mathcal{A}$, etc) needed in the ILP model.

\subsection{Determining visit duration \isr{t}{visit}}
\label{subsec:t-visit}

The basic building block in \AstroQ is a visit that requires \isr{t}{visit} slots. For KPF we compute \isr{t}{visit} via

\begin{equation}
\isr{t}{visit,r} = 
\frac{
  \isr{n}{exp,r} \isr{t}{exp,r} 
+ (\isr{n}{exp,r} - 1) \isr{t}{read} 
+ \isr{t}{slew}} 
{\isr{\tau}{slot}},
\label{eqn:slotsNeeded}
\end{equation}
where \isr{n}{exp,r} is the number of exposures per visit, \isr{t}{exp,r} is the exposure time, \isr{t}{read} is the CCD readout time (45~sec), and \isr{t}{slew} is the average slew \& acquisition time (120~sec). When $\isr{n}{exp,r} = 1$, CCD readout and slew occur concurrently and we do not budget for readout. We round \isr{t}{visit,r} to the nearest integer. 

Note that we budget for uniform target-to-target slew overheads. In reality, the time to slew between targets depends on their proximity and telescope cable wrap considerations. For a given night of observations, after we determine the optimal semester-level schedule, we run a second optimization, described in \cite{Handley2024_TTP}, to determine the optimal tour between targets on that night. At Keck, this reduces slew overheads by a factor of 5 over a random ordering. The time savings are used to slot-in bonus observations, if possible.

\subsection{Weather Accounting}
\label{Weather}

Even an initially optimal schedule is doomed to obsolescence due to unpredictable weather losses. To forecast realistic completion rates, \AstroQ must account for the certainty of weather losses.

Keck Observatory has tracked daily weather losses since 2002. We assembled a 22-year dataset spanning 2002-01-01 and 2023-12-31 that contains the time intervals when science observations (on any instrument) did not occur due to weather. Compared to other Keck instruments, Doppler programs are relatively resilient to poor weather since they contain bright stars ($V < 6$~mag) that can be observed through poor seeing or high extinction. Thus, we expect our historical dataset to be a slightly pessimistic view of expected KPF weather losses.

\begin{figure}
\centering\includegraphics[width=0.45\textwidth]{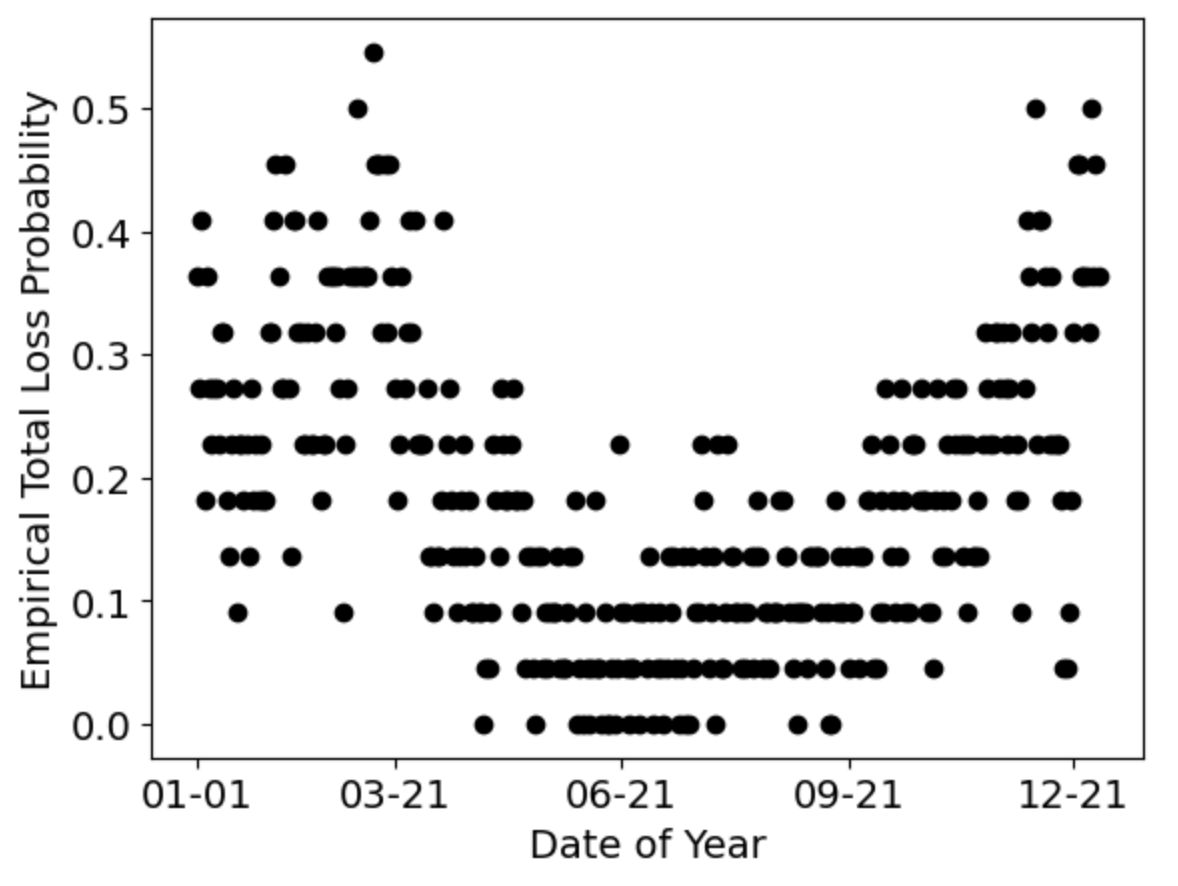}
    \caption{The fraction of times each calendar date was totally lost in our 22 year data set. The seasonal variations from winter to summer months are evident. A table containing the  information on each night's empirical total loss probability is available at the \AstroQ GitHub repository.}
    \label{fig:timelost}
\end{figure}

The distribution of lost time is bimodal: 33\% of all nights lose less than 25\% of their time to weather and 38\% of all nights lose more than 75\% of their time to weather. We approximate weather losses as negligible/complete if they are below/above 5 hours. Figure~\ref{fig:timelost} shows the fraction of nights with `total losses' averaged over the 22-year dataset.%
\footnote{The data behind this figure is included in the code repository}
This dataset shows a clear seasonality to weather losses: they peak in March and are lowest in July. Some calendar days in June/July have never been weathered out in the 22-year dataset. 

Additionally, weather losses are correlated between nights and we wish to account for the increased probability of adjacent nights to be co-lost. Based on the historical data, we found that if one night is lost there is a 14\% boost in the probability that the next night will be lost. 

Each time we run \AstroQ, we simulate a distribution of weather losses by:

\begin{enumerate}
\item Determine if $d$ = 1 is weathered out by flipping a coin weighted by the loss fraction shown in Figure~\ref{fig:timelost}.
\item If night $d$ is not weathered out, repeat weighted coin toss for $d + 1$.
\item If $d$ is weathered out, repeat weighted coin toss for $d + 1 $, but add 0.14 to the weight. 
\item Repeat steps (2-3) for all nights of the semester.
\end{enumerate}

We remove the weathered out nights from the allocation map before solving the semester. Whatever simulated weather losses, in our daily operations we assume fair weather for the upcoming night to produce a full observing plan.

As we conclude this section, we note that we are not attempting to construct a rigorous meteorological model for Maunakea. Rather, we built a simple empirical model of past patterns to ensure our forecasted program completion rates are not overly optimistic in the face of certain weather losses.

\section{Performance}
\label{sec:performance}

In this section, we benchmark \AstroQ against a number of observing strategies inspired by recent Doppler campaigns at the Keck Observatory. We define a nominal `request set' in \S\ref{sec:nominal}. We first schedule this request set into a real Keck allocation in \S\ref{sec:2018B}, noting {\em AstroQ's} computational needs and program completion rates. We repeat this experiment in \S\ref{sec:inst_allocation} and \S\ref{sec:inst_allocation_everynight} for different instrument allocations. In \S\ref{sec:weather-loss}, we consider weather losses, and in \S\ref{sec:test-single-shot} we compare the computational resources needed to schedule request sets with more short visits vs. fewer long visits. 

\subsection{Nominal Request Set}
\label{sec:nominal}

\input{toy_programs}

We defined a nominal request set comprised of six programs with a mixture of cadence requirements, total number of observations, number of targets, and number of intra-night visits. We adopted \isr{\tau}{\slot} = 5 minutes. We describe each program below and provide an example of a prior Doppler campaign at Keck Observatory that used a similar strategy:
\begin{itemize}
\item {\em Program 1---high cadence, nightly observations}. Strategy used in blind searches for low-mass, short-period planets around nearby stars; e.g., \cite{Howard2010}.

\item {\em Program 2---medium cadence, observe every third night}. Strategy used to measure masses of planets discovered by NASA's K2 mission with periods of weeks to months; e.g., \cite{Sinukoff2018}.

\item {\em Program 3---low cadence, observe every ten nights}. Strategy used in blind searches for distant giant planets with periods of months to years; e.g., \citealt{VanZandt2025}.

\item {\em Program 4---high cadence, multiple visits per night}. Strategy used to measure masses of ultra-short period planets, where significant acceleration is detectable over a single night; e.g., \cite{Dai2024}. 

\item {\em Program 5---high cadence, restricted coordinates}. Strategy used to measure masses of small planets from NASA's Kepler mission; e.g., \cite{Marcy2014}.

\item {\em Program 6---single shot}. Strategy used to spectroscopically characterize a large sample of planet-hosting stars from the Kepler mission; e.g., \cite{Petigura2017}.
\end{itemize}
Full details are listed in Table~\ref{tab:toy-strategies}.

\begin{figure*}
\centering\includegraphics[width=0.92\textwidth]{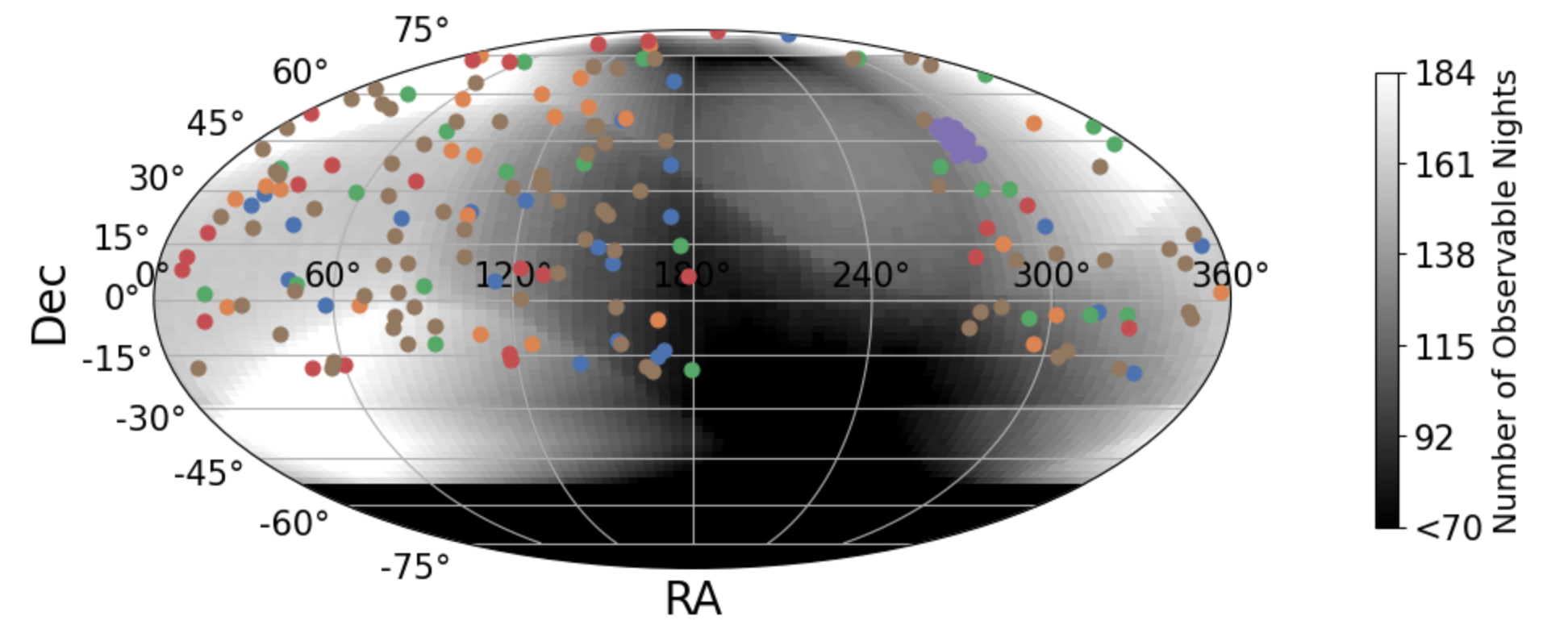}
\caption{Distribution of sky coordinates from the nominal request set described in \S\ref{sec:nominal}. Marker colors are the same on Figure \ref{fig:results-nominal}. The shading shows the number of unique B semester nights a target is observable as a function of RA and Dec. The dark-cap below Dec $-$50~deg are southern targets that never clear the minimum elevation requirements, the dark region near the North celestial pole circumpolar stars are often obstructed by the Keck-I Nasymth platform, the wave-shaped band is due to the Moon, the dark vertical band at $\sim$240~deg corresponds to RAs that are anti-aligned with the B semseter.}
\label{fig:nominal_allsky}
\end{figure*}

While the six programs were inspired by real observing campaigns at Keck Observatory, we standardized distribution of target coordinates and program time balance to ease interpretation of our benchmarks. For all but one program, we drew celestial coordinates RA $\alpha$ uniformly, excluding angles between 180~deg and 270~deg, and declination $\delta$ randomly between $-$20~deg and +90~deg following a uniform distribution in $\cos \delta$. This procedure produces a uniform distribution of targets per solid angle, as can be seen in Figure~\ref{fig:nominal_allsky}. In these experiments, we specified a seed for the random draws to ensure reproducibility. For Program 5, we drew coordinates to approximate the {\em Kepler} field $\alpha$ = 285--295~deg, $\delta$ = +40--50~deg.

Target coordinates generally favor a Keck B semester, but there is, of course, a range of accessibility as shown in the heatmap in Figure~\ref{fig:nominal_allsky}. Some stars are accessible for a portion of all 184 unique nights, e.g. `Star0014' at $(\alpha,\delta)$ = (352.4, +85.7)~deg; others are only accessible for 70 unique nights, e.g., `Star0068' at (179.8,$-$19.2)~deg.

All programs included 24 stars except for Program 6, which included 80. For each program, we selected \isr{t}{visit,r} such that $N_r \isr{n}{inter,r} \isr{n}{intra,max,r} \isr{t}{visit,r} = 960$~slots or 80 hours to achieve all requested visits. In total, the request set has 200 requests and 100\% completion would require 3680 visits, 5760 total slots, and 480 hours.

\newcounter{test}
\setcounter{test}{0}
\refstepcounter{test}
\label{test:2018B}
\subsection{Test \arabic{test}: Keck 2018B Allocation}
\label{sec:2018B}

For our first experiment, we adopted the 2018B Keck/HIRES schedule (with minor modifications) to serve as a realistic allocation of Doppler resources at Keck. During this time, HIRESr was scheduled for science operations on 173 quarter nights over 46 unique nights. We then supplemented this with 9 quarters across four nights to bring the total allocation to 182 quarters across 50 nights. During a B semester, the nights range from 9.4 to 12.0 hours in length and our total allocation is 481.6 hours or 5780 slots. If our requests had zero accessibility or cadence constraints, this allocation would result in 100\% completion. We are therefore free to interpret any shortfall to request/allocation mismatch or a failure of our optimizer to find an optimal solution.

We implemented the full-semester ILP model in Python and translated it into the Gurobi optimization language the Gurobi-py API \citep{gurobi}. We ran the model on an Apple Mac Studio with a 24-core M2 Ultra chip and 128GB of RAM. The constraint matrix had a total of \param{ns12-start_rows} rows (one per constraint), \param{ns12-start_cols} columns (one per decision variable), and \param{ns12-start_nonzeros} non-zero entries. Gurobi presolve reduced these values to \param{ns12-presolved_rows}, \param{ns12-presolved_cols}, and \param{ns12-presolved_nonzeros}, respectively. We instructed Gurobi to terminate when the integer feasible solution is within 1\% of the relaxed LP bound; it did so when the gap was \param{ns12-gap}\%, i.e. an optimal solution. Model construction, pre-solve, and solve took \param{ns12-build_time}, \param{ns12-presolve_time}, \param{ns12-solver_time} seconds respectively and required $\sim$1.25 GB of RAM. The best incumbent was Obj = \param{ns12-best_incobj}, or a total shortfall of \param{ns12-best_incobj} slots. This is equal to 9.6\% of the total requests, $\sim$46 hours, or $\sim$3.5 nights. Even in this engineered example, there is a mismatch between request and allocation. This strongly motivates a more sophisticated method for matching the nights allocated to the queue to the given set of requests. It further highlights how easy it is to make an infeasible request. 

\begin{figure*}
\vspace{-1cm}
\centering\includegraphics[width=0.90\textwidth]{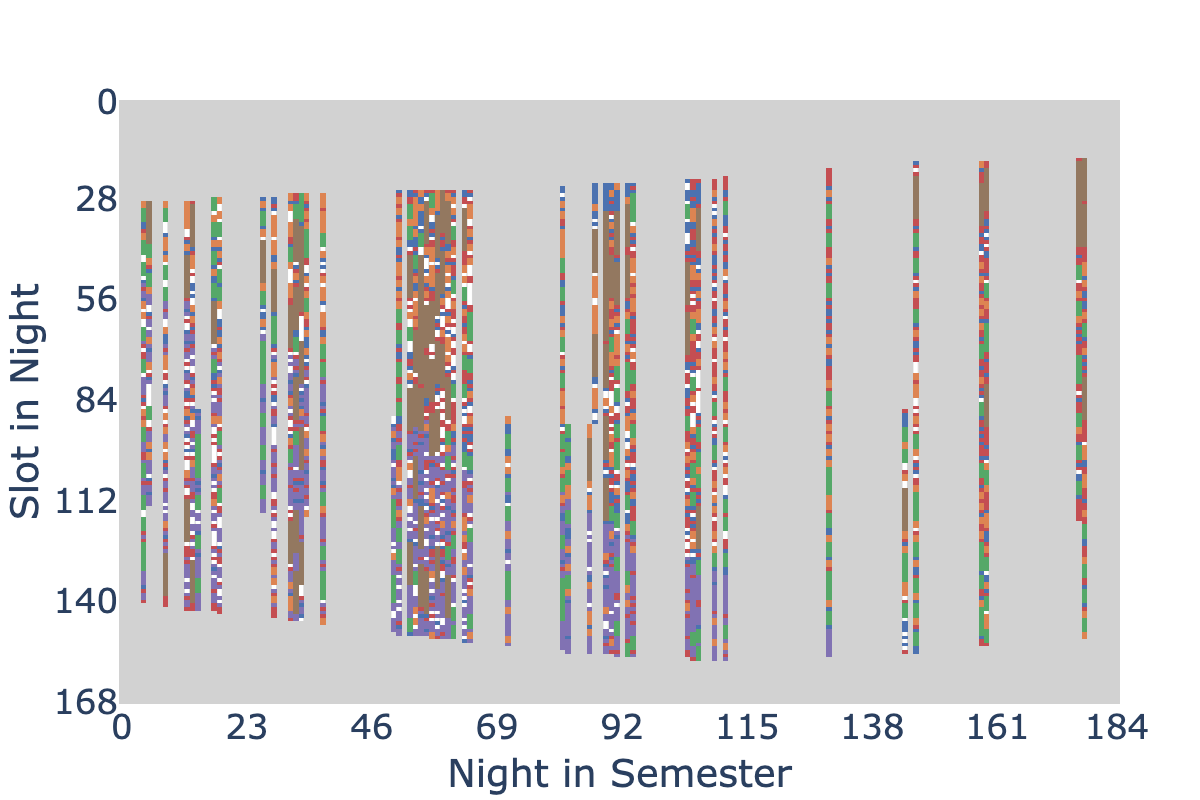}
\centering\includegraphics[width=0.90\textwidth]{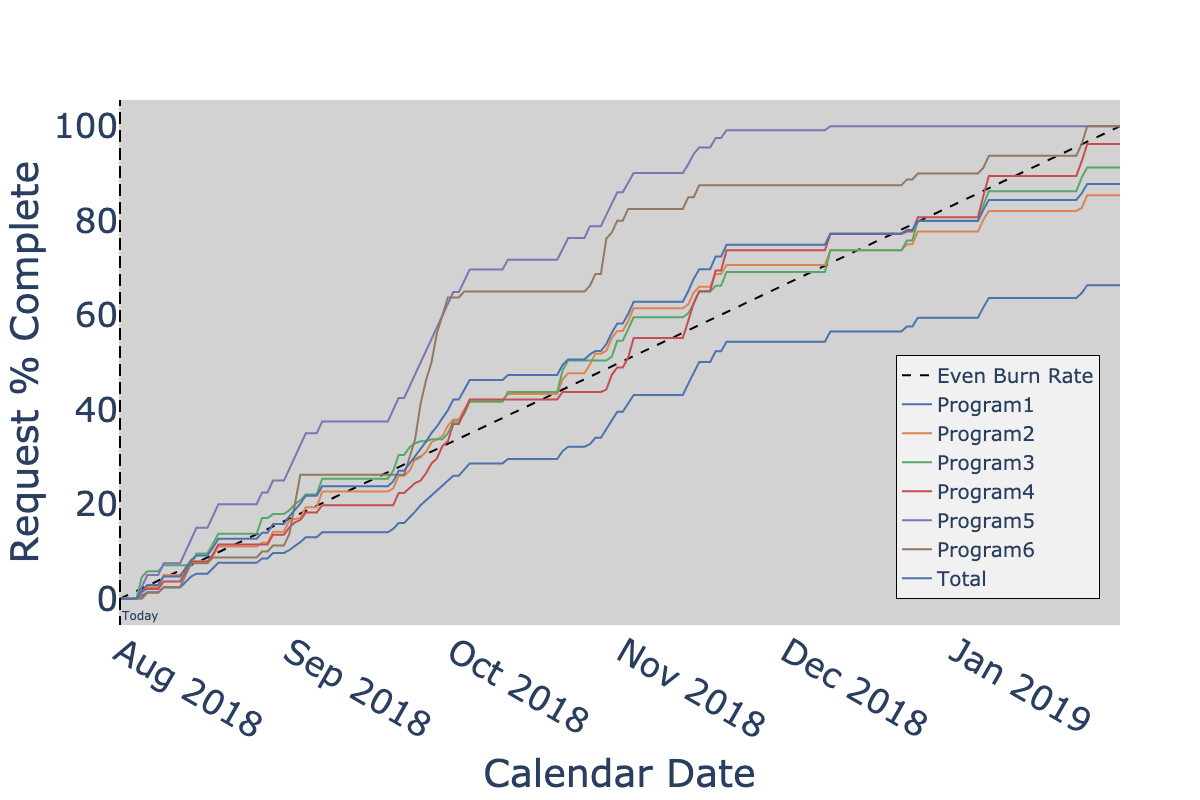}
\caption{{\bf Top: Bird's eye view of the optimal semester schedule described in \S\ref{sec:2018B}}. The plot is analogous to Figure~\ref{fig:slot-grid} with visits that are color-coded by program (key on bottom plot). Some slots are not filled despite this being the optimal solution. This is due to stars with infeasible requests due to challenging RAs. {\bf Bottom: The Cumulative Observation Function (COF)}. Each line shows the program completion over time. We comment on the dispersion in forecasted completion in \S\ref{sec:2018B}.}
\label{fig:results-nominal}
\end{figure*}

Figure~\ref{fig:results-nominal} shows two views of the solution described above. The top panel shows the filled/empty slots using the same axes as Figure~\ref{fig:slot-grid}. The small fraction of white slots indicates a nearly full schedule. In this toy model, a more careful consideration of target coordinates could have better filled these slots.  Recall that this schedule is optimal; it is simply not possible to schedule more observations into the allocated time. This occurred even though total requested slots were slightly less than the total allocated slots. The shortfall is due to a mismatch between the request set and available slots. As we will show later, some of this shortfall can be reduced by a different allocation of telescope time.

The bottom panel of Figure~\ref{fig:results-nominal} shows program completion percentage over time or `cumulative observation function' or COF. Programs 5 and 6 achieved \param{NS12only-prog6_complete}\% completion. It is easy to schedule Program 5 `Kepler field' because the allocation distribution favored nights early in the semester when the Kepler field is accessible (August through November). Program 6 consists of independent single-shot observations which are easy to schedule. The COFs of Programs 2, 3, and 4 follow the one-to-one line, as expected for targets with a uniform distribution of RA.

Program 1 has the lowest forecasted completion at \param{NS12only-prog1_complete}\%. The 24 requests in this program specified single visits on 40 unique nights but with only 50 unique nights on sky, there is little margin for error. Some stars in this program are not accessible every night. We confirmed each star with $<60\%$ individual completion rate (5 in total) had $\alpha$~=~145--180~deg, near the edge of our allowed range and at hours that are unobservable at the beginning of the semester when the density of nights is greatest. That said, 5 additional stars in Program 1 also fell into this RA range and completed $>60\%$. In contrast, Program 5 had the same observing strategy, but with stars confined to favorable RAs. 

Programs 2 and 3 had the second and third lowest completion rates at \param{NS12only-prog2_complete}\% and \param{NS12only-prog3_complete}\%, respectively. These corresponded to the medium-cadence ($\isr{\tau}{inter} = 3$~days) and low-cadence strategy ($\isr{\tau}{inter} = 10$~days). In Program 2, to achieve 20 observations every 3 days at best requires the target to be observable for a minimum of 58 days in the 184 day semester, for Program 3 to achieve 10 observations every 10 days requires at least 89 days of accessibility. For stars with unfavorable RAs achieving $\isr{\tau}{inter,max}$ is infeasible. 

Through our toy model, we understand the disparity in program completion rates. In the following sections, we will compare this nominal schedule to two schedules resulting from two alternate allocation selection functions. Through these tests, we show how program completion is heavily dependent on the distribution of allocated nights, and how different distributions favor or disfavor certain observing strategies.

\refstepcounter{test}
\label{test:2018B-random}
\subsection{Test \arabic{test}: Random Permutations of Keck 2018B Allocation}
\label{sec:inst_allocation}
In the previous section, we conjectured that some of the shortfall was due to request/allocation mismatch. We investigated this by rearranging time into alternate instrument allocations. We first generated 10 random time allocations by re-assigning the Test~1 allocations to different nights. We scheduled the same request set into each realization and report the mean and standard deviation of each program's completion percentage in Table~\ref{tab:toy-strategies}. The best objective ranged from \param{rand-all_min_incobj} to \param{rand-all_max_incobj}; as before, the optimality gap was less than 1\%. 

The average completion rate of Test 2 was \param{all_Random_avg_completion}\%, similar to that of Test 1 at \param{all_NS_avg_completion}\% . However, different programs are favored in the two tests and in some iterations of this test, the best incumbent objective can schedule $\sim$100 additional slots than the nominal request set test. Program 5, which completed in Test 1, suffers with randomized allocations since they do no favor its RA range. A more even distribution of nights is more favorable for Program 1 with its uniform distribution of RA. Meanwhile, Programs 2, 3, and 4 now enjoy more nights where their targets are available and so they nearly complete in every iteration. Note that the dispersion on completion rate is small (only a few percent) for all programs except Program 5, which drives the spread in the best incumbent objective across the runs. 

\refstepcounter{test}
\label{test:quarter}
\subsection{Test \arabic{test}: Quarter-night Allocations}
\label{sec:inst_allocation_everynight}

Next, we investigated distributing the same amount of time into only quarter nights to achieve the maximum number of nights on sky. We generated 10 random allocations where 182 quarter nights are scheduled into 184 unique nights. We required a 50/50 split between first/last quarters.

\AstroQ finds solutions with <1\% optimality gaps in a similar amount of time as in Tests 1 \& 2. The average program completion is higher than in Test 1 for every program except Program 4, the multi-visit program. This program suffers because the telescope is allocated no more than 3 hours in a given night. This limits the number of observations on given night to the accepted minimum of 3 rather than the desired maximum of 5, severely restricting this program. 

\refstepcounter{test}
\label{test:weather}
\subsection{Test \arabic{test}. Weather Losses and Program Completion}
\label{sec:weather-loss}
Next, we tested {\em AstroQ's} sensitivity to weather losses. We took the Test 1 allocation and weathered out nights, according to the procedure described in $\S$\ref{Weather} on 10 iterations, each with a different random seed. The mean and standard deviation of total lost time in these iterations was 19 $\pm$ 5\%. Interestingly, for some realizations, Gurobi had more difficulty optimizing this over-full schedule and terminated when it exceeded its wall-clock limit of 60~min, before the gap had dipped below the 1\% termination criterion. The best incumbent objective varied between \param{weath-all_min_incobj} and \param{weath-all_max_incobj} and completion was mostly with gaps less than 5\%, with two iterations as high as 20\%. Clearly, the optimization problem is harder for under-allocated semesters. 

Table~\ref{tab:toy-strategies} summarizes the completion rates as before. The total average completion across the weather simulations falls to \param{all_Weather_avg_completion}\% $\pm$ \param{all_Weather_avg_deviation}\%, about $\sim$11\% lower than the lossless test. This fractional drop is smaller than the average fraction of lost time. \AstroQ is able to put some of the unscheduled time in Test~\ref{test:2018B} toward making up for weather losses.

Some programs are more resilient to weather losses than others. Program 1 and 5 are very sensitive to weather loss since they require nearly nightly observations. On the other hand, Program 6 is flexible and insensitive to weather losses. Meanwhile Programs 2 and 3 are minimally affected by weather since they require fewer on-sky nights. Lastly, Program 4 suffers with large deviation. Similar to Test \ref{test:quarter}, this is because in order to schedule at multiple visits, there must be a long enough period in the night where both the target is at high enough elevation to observe and the telescope is allocated to the queue. This is easiest on full nights; but when weather losses randomly fall on these full nights more so than partial nights, multi-visit stars are disproportionally affected.

\subsection{Number of Single Shot Exposures}
\label{sec:test-single-shot}

Finally, we explored the computational complexity of \AstroQ while trading more short exposures for fewer long exposures. Starting with Test~\ref{test:2018B}, we decreased the \isr{t}{visit} for Program 6 from 12, to 6, to 3, to 2, to 1 slot and increased the number of stars from 80, to 160, to 320, to 480, to 960. At all steps, the optimizer found a suitable solution with an optimality gap of < 1\% in $\lesssim200$~sec, indicating the model complexity is not largely driven by the number of multi-slot requests.

\section{Conclusion}
\label{sec:conclusion}
In this paper, we have introduced the \AstroQ auto-scheduling algorithm and have applied it to the Doppler needs of KPF at Keck Observatory. \AstroQ can achieve a globally optimal solution for 200 targets --- each with their own cadence needs and accessibility constraints --- over a six-month semester at five-minute time resolution in a few minutes on a modern workstation. This speed means \AstroQ is well-suited for dynamic scheduling applications and can respond to changing weather conditions, target-of-opportunity interrupts, or other demands. It is a tool that can reduce both the human effort needed to schedule observations and the potential for bias in executing portfolio of observing programs. 

In this work, we considered the problem of maximizing program completion subject to a single allocation distribution. One could apply the same formalism outlined in this paper toward determining the optimal instrument allocation. Here, one would attempt to maximize program completion over many possible allocations subject to various constraints such as total awarded time to the queue, blackout dates, and other operational needs of the observatory. 

In closing, we note that while \AstroQ was developed for the PRV needs of KPF users at Keck Observatory, it is agnostic to the observatory or science domain. We hope \AstroQ may be adopted or adapted for different science needs and observatories. 

\section{Acknowledgments}
\label{Acknowledgements}

We recognize and acknowledge the cultural role and reverence that the summit of Maunakea has within the indigenous Hawaiian community. We are deeply grateful to have the opportunity to conduct observations from this mountain. 

J.L, E.P., J.V.Z. are grateful to the Heising-Simons Foundation supporting this work under grant numbers 2022-3832 \& 2022-3833. 

The authors are grateful to all the past, present, and future observers, who are mostly graduate students that have, currently do, and/or will donate so much of their professional and personal time to staying up through all hours of the night to execute observations on behalf of many PIs within the framework of the queue. They were instrumental in testing and providing feedback on \AstroQ. 

We acknowledge extraordinary contributions of Keck Observatory staff who have helped advise, support, guide, and implement the KPF-Community Cadence project, which motivated this work. These people include hardware and software engineers, technicians, Observing Assistants, Staff Astronomers, and administrators who make all this science possible. We wish to especially recognize Jeff Mader and Tyler Coda from the Keck software group, Carolyn Jordan the Keck Scheduler and Observing Assistant Manager, and Josh Walawender the KPF Lead Staff Astronomer. 

\appendix
\section{Symbols used in this work}
\input{variablesTable}

\bibliography{bib.bib}

\end{document}

%% file: params.tex
\usepackage{xparse} 
\usepackage{xcolor} 
\usepackage{etoolbox} 
\ExplSyntaxOn 
\NewDocumentCommand{\param}{m}{ 
\str_case:nnF {#1} { 
{ns12-build_time}{15}%
{ns12-presolve_time}{21}%
{ns12-solver_time}{172}%
{ns12-total_time}{194}%
{ns12-start_rows}{67965}%
{ns12-start_cols}{365924}%
{ns12-start_nonzeros}{3986346}%
{ns12-presolved_rows}{50013}%
{ns12-presolved_cols}{361937}%
{ns12-presolved_nonzeros}{3211022}%
{ns12-best_incobj}{554}%
{ns12-best_bound}{554}%
{ns12-gap}{0}%
{ns12-sum_theta}{421}%
{ns6-build_time}{17}%
{ns6-presolve_time}{26}%
{ns6-solver_time}{134}%
{ns6-total_time}{160}%
{ns6-start_rows}{71063}%
{ns6-start_cols}{525583}%
{ns6-start_nonzeros}{4682773}%
{ns6-presolved_rows}{50094}%
{ns6-presolved_cols}{521597}%
{ns6-presolved_nonzeros}{3586983}%
{ns6-best_incobj}{556}%
{ns6-best_bound}{554}%
{ns6-gap}{0}%
{ns6-sum_theta}{423}%
{ns3-build_time}{21}%
{ns3-presolve_time}{35}%
{ns3-solver_time}{131}%
{ns3-total_time}{166}%
{ns3-start_rows}{77337}%
{ns3-start_cols}{867104}%
{ns3-start_nonzeros}{5882550}%
{ns3-presolved_rows}{50253}%
{ns3-presolved_cols}{863142}%
{ns3-presolved_nonzeros}{4103956}%
{ns3-best_incobj}{555}%
{ns3-best_bound}{554}%
{ns3-gap}{0}%
{ns3-sum_theta}{422}%
{ns2-build_time}{25}%
{ns2-presolve_time}{32}%
{ns2-solver_time}{117}%
{ns2-total_time}{161}%
{ns2-start_rows}{83620}%
{ns2-start_cols}{1207559}%
{ns2-start_nonzeros}{6960081}%
{ns2-presolved_rows}{50412}%
{ns2-presolved_cols}{1203571}%
{ns2-presolved_nonzeros}{4499889}%
{ns2-best_incobj}{558}%
{ns2-best_bound}{554}%
{ns2-gap}{0}%
{ns2-sum_theta}{422}%
{ns1-build_time}{36}%
{ns1-presolve_time}{42}%
{ns1-solver_time}{122}%
{ns1-total_time}{194}%
{ns1-start_rows}{102157}%
{ns1-start_cols}{2216374}%
{ns1-start_nonzeros}{10029822}%
{ns1-presolved_rows}{51481}%
{ns1-presolved_cols}{2212385}%
{ns1-presolved_nonzeros}{5556394}%
{ns1-best_incobj}{554}%
{ns1-best_bound}{554}%
{ns1-gap}{0}%
{ns1-sum_theta}{421}%
{ns-all_max_incobj}{558}%
{ns-all_min_incobj}{554}%
{rand-all_max_incobj}{628}%
{rand-all_min_incobj}{400}%
{every-all_max_incobj}{136}%
{every-all_min_incobj}{100}%
{weath-all_max_incobj}{1293}%
{weath-all_min_incobj}{714}%
{NS12only-prog1_complete}{66}%
{NS12only-prog1_deviation}{0.0}%
{NS12only-prog2_complete}{85}%
{NS12only-prog2_deviation}{0.0}%
{NS12only-prog3_complete}{91}%
{NS12only-prog3_deviation}{0.0}%
{NS12only-prog4_complete}{96}%
{NS12only-prog4_deviation}{0.0}%
{NS12only-prog5_complete}{100}%
{NS12only-prog5_deviation}{0.0}%
{NS12only-prog6_complete}{100}%
{NS12only-prog6_deviation}{0.0}%
{all_NS12only_avg_completion}{90}%
{all_NS12only_avg_deviation}{0.0}%
{NS-prog1_complete}{66}%
{NS-prog1_deviation}{0.0}%
{NS-prog2_complete}{85}%
{NS-prog2_deviation}{0.0}%
{NS-prog3_complete}{91}%
{NS-prog3_deviation}{0.2}%
{NS-prog4_complete}{96}%
{NS-prog4_deviation}{0.3}%
{NS-prog5_complete}{99}%
{NS-prog5_deviation}{0.0}%
{NS-prog6_complete}{100}%
{NS-prog6_deviation}{0.0}%
{all_NS_avg_completion}{90}%
{all_NS_avg_deviation}{0.1}%
{Random-prog1_complete}{80}%
{Random-prog1_deviation}{2.5}%
{Random-prog2_complete}{96}%
{Random-prog2_deviation}{0.9}%
{Random-prog3_complete}{93}%
{Random-prog3_deviation}{1.3}%
{Random-prog4_complete}{98}%
{Random-prog4_deviation}{0.8}%
{Random-prog5_complete}{76}%
{Random-prog5_deviation}{6.7}%
{Random-prog6_complete}{100}%
{Random-prog6_deviation}{0.0}%
{all_Random_avg_completion}{91}%
{all_Random_avg_deviation}{2.0}%
{EveryNight-prog1_complete}{99}%
{EveryNight-prog1_deviation}{0.8}%
{EveryNight-prog2_complete}{99}%
{EveryNight-prog2_deviation}{0.1}%
{EveryNight-prog3_complete}{97}%
{EveryNight-prog3_deviation}{0.5}%
{EveryNight-prog4_complete}{59}%
{EveryNight-prog4_deviation}{0.1}%
{EveryNight-prog5_complete}{100}%
{EveryNight-prog5_deviation}{0.0}%
{EveryNight-prog6_complete}{100}%
{EveryNight-prog6_deviation}{0.0}%
{all_EveryNight_avg_completion}{93}%
{all_EveryNight_avg_deviation}{0.2}%
{Weather-prog1_complete}{51}%
{Weather-prog1_deviation}{5.2}%
{Weather-prog2_complete}{76}%
{Weather-prog2_deviation}{5.2}%
{Weather-prog3_complete}{84}%
{Weather-prog3_deviation}{4.1}%
{Weather-prog4_complete}{72}%
{Weather-prog4_deviation}{13.9}%
{Weather-prog5_complete}{87}%
{Weather-prog5_deviation}{5.6}%
{Weather-prog6_complete}{99}%
{Weather-prog6_deviation}{0.5}%
{all_Weather_avg_completion}{79}%
{all_Weather_avg_deviation}{5.7}%
{all_rand_avg_bestincobj}{525}%
{all_rand_dev_bestincobj}{63}%
{all_every_avg_bestincobj}{109}%
{all_every_dev_bestincobj}{11}%
{all_weath_avg_bestincobj}{1012}%
{all_weath_dev_bestincobj}{197}%
}{{\color{red}XX}}  
} 
\ExplSyntaxOff

%% file: toy_programs.tex
\begin{table*}[t]
  \centering
  \caption{Description of Toy Programs}
  \label{tab:toy-strategies}
  \begin{tabular}{@{}lrrrrrrrrrr@{}}
    \toprule
    Prog. & $N_r$ & $n_{\mathrm{inter,r}}$ & $\tau_{\mathrm{inter,r}}$ & $n_{\mathrm{intra,r}}$ & $t_{\mathrm{visit}}$ & Total time & \multicolumn{4}{c}{Completion fraction (\%)} \\
    \cmidrule(lr){8-11}
    & & & (days) & max/min & (slots) & (slots) & Test~\ref{test:2018B} & Test~\ref{test:2018B-random} & Test~\ref{test:quarter}&Test~\ref{test:weather}  \\
    \midrule
    1 & 24 & 40 & 1  & 1/1     & 1  & 960  & \param{NS12only-prog1_complete} & \param{Random-prog1_complete} $\pm$ \param{Random-prog1_deviation} & \param{EveryNight-prog1_complete} $\pm$ \param{EveryNight-prog1_deviation} & \param{Weather-prog1_complete} $\pm$ \param{Weather-prog1_deviation}  \\
    2 & 24 & 20 & 3  & 1/1     & 2  & 960  & \param{NS12only-prog2_complete} & \param{Random-prog2_complete} $\pm$ \param{Random-prog2_deviation} & \param{EveryNight-prog2_complete} $\pm$ \param{EveryNight-prog2_deviation} & \param{Weather-prog2_complete} $\pm$ \param{Weather-prog2_deviation}  \\
    3 & 24 & 10 &10  & 1/1     & 4  & 960  & \param{NS12only-prog3_complete} & \param{Random-prog3_complete} $\pm$ \param{Random-prog3_deviation} & \param{EveryNight-prog3_complete} $\pm$ \param{EveryNight-prog3_deviation} & \param{Weather-prog3_complete} $\pm$ \param{Weather-prog3_deviation}  \\
    4 & 24 & 8  & 1  & 5/3 & 1  & 960  & \param{NS12only-prog4_complete} & \param{Random-prog4_complete} $\pm$ \param{Random-prog4_deviation} & \param{EveryNight-prog4_complete} $\pm$ \param{EveryNight-prog4_deviation} & \param{Weather-prog4_complete} $\pm$ \param{Weather-prog4_deviation} \\
    5 & 24 & 40 & 1  & 1/1     & 1 & 960  & \param{NS12only-prog5_complete} & \param{Random-prog5_complete} $\pm$ \param{Random-prog5_deviation} & \param{EveryNight-prog5_complete} $\pm$ \param{EveryNight-prog5_deviation} & \param{Weather-prog5_complete} $\pm$ \param{Weather-prog5_deviation}  \\
    6 & 80 & 1  & 1  & 1/1     & 12  & 960   & \param{NS12only-prog6_complete} & \param{Random-prog6_complete} $\pm$ \param{Random-prog6_deviation} & \param{EveryNight-prog6_complete} $\pm$ \param{EveryNight-prog6_deviation} & \param{Weather-prog6_complete} $\pm$ \param{Weather-prog6_deviation}  \\
    \midrule
    Sum & 200 & & & & & 5760 & & & & \\
    \hline
    Avg Completion &     & & & & &      
    & \param{all_NS_avg_completion} 
    & \param{all_Random_avg_completion} $\pm$ \param{all_Random_avg_deviation} 
    & \param{all_EveryNight_avg_completion} $\pm$ \param{all_EveryNight_avg_deviation}
    & \param{all_Weather_avg_completion} $\pm$ \param{all_Weather_avg_deviation}  \\
    \hline
    Avg Incumbent &     & & & & &      
    & \param{ns12-best_incobj} 
    & \param{all_rand_avg_bestincobj} $\pm$ \param{all_rand_dev_bestincobj}
    & \param{all_every_avg_bestincobj} $\pm$ \param{all_every_dev_bestincobj}
    & \param{all_weath_avg_bestincobj} $\pm$ \param{all_weath_dev_bestincobj} \\
    \bottomrule
  \end{tabular}
  \noindent\parbox{\textwidth}{\footnotesize{$^a$\isr{\tau}{slot} = 5~min.}}
  \noindent\parbox{\textwidth}{\footnotesize{$^b$For Program 4, $\tau_{\mathrm{intra,r}} = 12$ slots or 1 hour}}
  \noindent\parbox{\textwidth}{\footnotesize{$^c$Test~\ref{test:2018B}---Real Keck 2018B allocation (\S\ref{sec:2018B}), Test~\ref{test:2018B-random}---Random permutation of Keck 2018B allocation (\S\ref{sec:inst_allocation}), Test~\ref{test:quarter}---all quarter night allocations (\S\ref{sec:inst_allocation_everynight}), Test~\ref{test:weather}---simulated weather losses (\S\ref{sec:weather-loss}}) .}
\end{table*}

%% file: variablesTable.tex
\begin{table*}[h]
  \centering
  \caption{Dictionary of Variables}
  \label{variableDict}
  \begin{tabular}{cl}
    \hline
    Notation & \multicolumn{1}{c}{Description} \\
    \hline
    \hline

    \hfill \\ 
    \multicolumn{2}{l}{\textbf{Indices:}} \\
    $r$ & a request  \\
    $s$ & a slot  \\
    $d$ & a night \\
    \hfill \\ 
    
    \multicolumn{2}{l}{\textbf{Grid Sizes:}} \\
    $N_r$ & Number of requests  \\
    $N_d$ & Number of nights in semester \\
    $N_s$ & Number of slots in a night  \\
    \hfill \\ 

    \multicolumn{2}{l}{\textbf{Sets:}} \\
    $\Rcal$ & Set of all requests \\
    $\Dcal$ & Set of all days \\
    $\Scal$ & Set of all slots \\ 
    $\mathcal{A}_r$ & $\{(d,s) \in \Dcal \times \Scal \mid \text{request $r$ is accessible in day-slot pair $(d,s)$} \}$ \\
    $\Acal$ & $\{(r, d,s) \in \Rcal \times \Dcal \times \Scal \mid \text{request $r$ is accessible in slot $s$ of day $d$} \}$ \\
    $\Rcal_{d,s}$ & $\{r \in \Rcal \mid \text{request $r$ is accessible in slot $s$ of day $d$} \}$ \\
    $\Scal_{r, d}$ & $\{s \in \Scal \mid \text{request $r$ is accessible in slot $s$ on day $d$} \}$ \\
    $\Dcal_{r}$ & $\{d \in \Dcal  \mid \text{request $r$ is accessible in at least one slot on day $d$} \}$  \\   
    $\Tcal$ & $\bigl\{ \displaystyle\bigcup_{r \in R} A_r \,\big|\, \text{day-slot pair $(d,s)$ has at least one request accessible} \bigr\}$ \\
    $\Rcal_{\multivisit}$ & $\{r \in \Rcal \mid \text{request $r$ requires more than 1 visit per night} \}$   \\
    $\Rcal_{\singlevisit}$ & $\{r \in \Rcal \mid \text{request $r$ requires exactly 1 visit per night} \}$     \\

    \hfill \\ 
    
    \multicolumn{2}{l}{\textbf{Observational Strategy:}} \\
    \isr{n}{intra, max, r} & Desired (maximum) number of visits per night for request $r$  \\
    \isr{n}{intra, min, r} & Accepted (minimum) number of visits per night for request $r$  \\
    \isr{\tau}{intra, r} & Minimum intra-night cadence for request $r$ \\   
    \isr{n}{inter, r} & Maximum number of desired visits for request $r$  \\
    \isr{\tau}{inter, r} & Minimum inter-night cadence for request $r$ \\
    \isr{t}{visit, r} & The number of slots required to complete a visit for request $r$  \\ 
    $P_r$ & The number of past nights where request $r$ was observed \\
    \hfill \\ 

    \multicolumn{2}{l}{\textbf{Matrix Variables:}} \\
    $Y_{r,d,s}$ & Matrix of scheduled visits  \\
    $W_{r,d}$ & Matrix of on-sky visits, only for $r \in \Rcal_{\multivisit}$  \\

    $\Theta_r$ & Matrix of shortfall \\
    \hfill \\ 
    
    \multicolumn{2}{l}{\textbf{Constants:}} \\
    $\tau_{\slot}$ & the time represented by one slot, in minutes \\

    \hline
  \end{tabular}
\end{table*}